\documentclass[floatfix,12pt,english]{iopart}
\usepackage[T1]{fontenc}
\usepackage[latin1]{inputenc}
\usepackage{float}
\usepackage{graphicx}
\usepackage{epsfig}
\usepackage{multicol}
\usepackage{latexsym}
\usepackage{amsfonts}
\usepackage{bbold}
\usepackage{graphicx}
\usepackage{babel}


\usepackage[T1]{fontenc} % fontes com acentos no PDF
\def\Box{\kern1pt\vbox{\hrule height 1.2pt\hbox{\vrule width 1.2pt\hskip 3pt 
\vbox{\vskip 6pt}\hskip 3pt\vrule width 0.6pt}\hrule height 0.6pt}\kern1pt} 
\def\be{\begin{equation}}
\def\ee{\end{equation}}

\def\be{\begin{equation}}
\def\ee{\end{equation}}
\def\bea{\begin{eqnarray}}
\def\eea{\end{eqnarray}}

\begin{document}

\title{Physical approximations for the nonlinear evolution of perturbations in 
dark energy scenarios}
% spherical collapse,
% pseudo-Newtonian or general relativistic?
%

\author{L. R. Abramo$^1$, R. C. Batista$^1$, L. Liberato$^2$
and R. Rosenfeld$^2$}

\address{$^1$ Instituto de F\'{\i}sica, Universidade de S\~ao Paulo\\
CP 66318, 05315-970, S\~ao Paulo, Brazil}
\address{$^2$ Instituto de F\'{\i}sica Te\'orica, Universidade Estadual Paulista\\
R. Pamplona 145, 01405-900, S\~ao Paulo, Brazil}

\eads{\mailto{abramo@fma.if.usp.br},
\mailto{rbatista@fma.if.usp.br},
\mailto{liberato@ift.unesp.br},
\mailto{rosenfel@ift.unesp.br}}

\begin{center}
\today
\end{center}

Cosmology: large-scale structure of the Universe

\begin{abstract}
The abundance and distribution of collapsed objects such as
galaxy clusters will become an important tool to investigate
the nature of dark energy and dark matter. Number counts
of very massive objects 
%($10^{14} - 10^{16} M_\odot$) 
are sensitive
not only to the equation of state of dark energy, which parametrizes
the smooth component of its pressure, but also to the sound speed of
dark energy as well, which determines the amount of pressure in
inhomogeneous and collapsed structures. Since the evolution of
these structures must be followed well into the nonlinear
regime, and a fully relativistic framework for this regime does
not exist yet, we compare two approximate schemes: the 
widely used spherical
collapse model, and the pseudo-Newtonian approach. 
We show that both approximation schemes convey identical equations
for the density contrast, when the pressure perturbation of dark 
energy is parametrized in
terms of an effective sound speed. We also make a comparison 
of these approximate approaches to general relativity in 
the linearized regime, which lends some support 
to the approximations.
%We indicate which
%situations are likely to be well described by these approximations, 
%and when they are likely to fail.
\end{abstract}

\maketitle

\section{Introduction}

We now have overwhelming evidence that the Universe is accelerating, possibly
under the influence of some type of negative-pressure substance --
dark energy (DE) \cite{Frieman:2008sn,Albrecht:2006um,Seljak:2006bg}.
However, even though DE may be directly responsible for this
enhanced expansion, it is widely believed that the direct impact of
perturbations in DE density and pressure 
on structure formation is very weak.
This is strictly correct only for a cosmological constant model of DE, which 
does not have perturbations. 

For most scalar field models of
DE, this component remains very homogeneous 
even on galaxy and cluster scales. 
Heuristically, this can be understood as follows. 
In these models the scalar field can not have relaxed
to its minimum energy state and one must require that the time scale 
for the variation of the field is longer than the Hubble time,
implying a very flat potential. 
Therefore, the scalar field must be
extraordinarily light,  $m < H_0$, where $H_0$ is the Hubble parameter today.
The mass of the scalar field sets the
scale for its spatial variation and hence one usually expects small perturbations in the scalar
field for scales $\lambda < 1/m$ (the Compton wavelength), 
which are of the order of the Hubble radius. 
However, this argument may not apply to more general models of dark energy.

%Hence, one must require that the time scale 
%for the variation of the field is longer than the Hubble time, $\dot{\phi}/\phi < H_0$,
%where $\phi$ denotes the scalar field, $H_0$ is the Hubble parameter today
%and $\dot{} = d/dt$. Combining this condition with the equation of motion for the
%scalar field in the slow-roll regime results in $1/\phi dV(\phi)/d\phi < H_0^2$, where
%$V(\phi)$ is the scalar field potential. For a quadratic potential of the form $V(\phi) =  m^2 \phi^2/2$
%this implies for the mass of the scalar field $m < H_0$. The mass of the scalar field sets the
%scale for its spatial variation and hence one usually expects small perturbations in the scalar
%field for scales $\lambda < 1/m$, which are of the order of the Hubble radius. 
%However, this argument may not apply to more general models of dark energy.

If our only concern is the evolution of the background, 
then the role of dark energy in the evolution of dark matter perturbations 
is completely determined by its 
equation of state $w=p_e/\rho_e$, where $p_e$ is the homogeneous
pressure and $\rho_e$ is the homogeneous energy density of dark energy
\cite{Peebles:2002gy,Padmanabhan:2003gd,Sahni:2004ai}. 
At this level, dark energy affects structure formation indirectly
because, as it starts to dominate the background, 
very large structures are ripped apart by the ensuing
accelerated expansion
\cite{Linder:2005in,Liberato:2006un}.

However, dark energy can influence structure formation in an
additional manner. If it is a dynamical field or fluid,
then dark energy must possess inhomogeneities, and these perturbations
will interact gravitationally both with themselves and with clumps of
dark matter \cite{Coble:1996te}. This means that, unless dark energy
is just a cosmological constant, it will both feel and
create local gravitational potentials. 

Although the effect of these inhomogeneities in the dark energy 
component becomes small as $w \rightarrow -1$, in many
models with $w \neq -1$ it can be non-negligible when evolved 
in the nonlinear regime \cite{Basilakos:2003bi, Klypin:2003ug, Linder:2003dr, Lokas:2003cj, Kuhlen:2004rw, Mota:2004pa, Nunes:2004wn, Nunes:2005fn, Horellou:2005qc, Manera:2005ct, Dutta:2006pn, Abramo:2007iu, Mota:2007zn, Abramo:2007mv}. 
Since the effects of dark energy perturbations on the cosmic microwave
background are quite small (see, e.g. \cite{Abramo:2004ji}),
structure formation is the only remaining probe of the nature 
of dark energy on small and intermediate scales. 

Nevertheless, a fully relativistic method to treat nonlinear 
perturbations is not available. 
When there is a pressure ingredient the 
nonlinear relativistic equations take a very complicated form.
The Lema\^itre-Tolman-Bondi (LTB) model
\cite{Lemaitre:1933gd,Tolman:1934za,Bondi:1947av} 
is the closest one can get to a working formalism,
but it only works if matter is pressureless.
The problem is not with the gravity side of the equations,
but with the nonlinear evolution of matter and the relativistic
treatment of pressure. 

In this respect, 
the only well-studied models with inhomogeneous dark energy are those involving 
canonical scalar fields 
\cite{Mota:2004pa, Nunes:2004wn, Nunes:2005fn, Horellou:2005qc, Manera:2005ct, Dutta:2006pn}, for which
the equations of motion and the pressure
%(the trace of the spatial components of the stress-energy tensor) 
follow directly from a given Lagrangian. For these models
the free parameters are the scalar potential and some set
of initial conditions. In this approach,
the equation of state, the density perturbations and the 
pressure perturbations are derived quantities.
Hence, a more kinematical  and model-independent approach to structure formation, 
closer in spirit to the homogeneous description of dark energy in 
terms of a parametrized equation of state $w(z)$, is sorely lacking.
% But since in that case the adiabatic 
% sound speed is $c_s^2 = 1$, canonical scalar fields are a 
% particular case where dark energy fluctuations are suppressed
% even if $w \neq -1$. Hence

%In previous works we used 
%the SC/PN approach to study the non-linear evolution of a system of 
%perfect fluids with pressure.
%In \cite{Abramo:2007iu} we considered a dark energy model where
%the effective sound speed was equal to the equation of state, $c_{\rm eff}^2 = w$,
%and we computed the number counts for that model.
%In \cite{Abramo:2007mv} we considered a more general effective sound speed,
%and we showed that the ``effective equation of state'' of dark energy inside
%a collapsed region could switch from that of a cosmological constant
%to that of cold dark matter.

There are two very different approximations to full-blown
general relativity that have been frequently used.
They are the spherical collapse (SC) model 
\cite{Gunn:1972sv, Padmanabhan, Fosalba:1997tn}
and the pseudo-Newtonian (PN) approach \cite{Lima:1996at, Hwang:1997xt, Reis:2003fs, Hwang:2005xt}. 
We have recently used these approximations in the nonlinear
regime in order to show that the ``effective equation of state'' of dark energy inside
a collapsed region could be very different from its background value \cite{Abramo:2007mv}.

In this work we show that, even though
the underlying assumptions for either approach are rather different,
they yield exactly the same nonperturbative equations as long as the pressure
perturbations are treated in the same way. They also have an 
important advantage: they allow for a completely parametrized approach
to dark energy. Furthermore, we compare the growth of perturbations
in the linear regime with a linearized relativistic analysis and show that they 
are similar, lending support to the approximations. 

This paper is organized as follows. 
In Section 2 we review both the PN and SC approaches and show that they are equivalent.
In Section 3 we study the linear evolution of perturbations in DE in this approximation.
The linear evolution of perturbations in a universe with a 2 component fluid is
studied in general relativity in Section 4. We present a comparison between the
relativistic analysis and the approximate analysis in the linear regime in Section 5.
Section 6 concludes.

\section{Spherical collapse and pseudo-Newtonian cosmology}

In linear perturbation theory there are essentially three
degrees of freedom for scalar perturbations:
the energy density perturbation $\delta\rho$, 
the pressure perturbation $\delta p$ and the
scalar anisotropic stress $\pi$ \cite{Bardeen:1980kt,Kodama:1985bj}.
An alternative set is given by the density contrast $\delta\rho/\rho$,  the velocity potential
$\theta=\vec\nabla \cdot \vec{v}$ and
the anisotropic stress \cite{Ma:1995ey}. Since large-scale 
anisotropic stresses decay rapidly, they can only become relevant 
again inside structures which have collapsed. This means that
anisotropic stress should not influence the mass of these 
structures, and therefore it is unlikely that dark energy models can
be differentiated on the basis of anisotropic stress. For this
reason we do not consider it any further in this work (see, however, \cite{Koivisto:2007bp}).

We will parametrize the pressure perturbation 
using the so-called effective sound velocity
\cite{Hu:1998kj}, defined as $c_{\rm eff}^2 \equiv \delta p_e/ \delta \rho_e$.
We will assume that $c_{\rm eff}^2$ is a function of time only,
even though this simplification lacks any formal basis in 
cosmological perturbation theory. This should be clear from the fact that
$\delta p_e$ is an independent degree of freedom whose time and 
spatial dependences can be, and often are, completely different from
$\delta\rho_e$.
Only in a particular gauge (the so-called ``rest frame''
of the fluid, where $T^i_0 = 0$) the effective sound speed
coincides with the universal sound speed of linear relativistic
perturbations, $c_X^2$ \cite{Hu:1998kj,Mukhanov:05}.
It may be difficult to realize this parametrization in a natural
model, but the situation is not much different from what happens
when we parametrize the equation of state.

%%%%%%%%%%%%%%%%%%%%%%%%%%%
%%
%% RONALDO (mutation2)
%%
%%%%%%%%%%%%%%%%%%%%%%%%%%%

The main reason that we use the effective sound speed, though, is
that it allows us to study nonlinear structure formation
within the spherical collapse model \cite{Gunn:1972sv}. 
In this extremely simple model, a spherically symmetric
region of homogeneous overdensity 
evolves inside the homogeneous expanding Universe 
(this is the so-called ``top-hat'' density profile). General relativistic
arguments show that one can regard the overdense region as
a mini-universe of positive curvature, and then we use the Friedmann 
and the Raychaudhury equations to evolve
the density and radius of the spherical region \cite{Padmanabhan, Fosalba:1997tn}.

It is therefore extremely interesting that this simplified relativistic
approach coincides with a pseudo-Newtonian approach to cosmology.
In fact, we will show below that, as long as the pressure perturbations
are described in terms of an effective pressure, the two approximations
are completely equivalent. This means that the
main physical characteristics of gravitational collapse of structures 
such as galaxy clusters is probably well described within this framework.

The argument is as follows. First, the SC approach should be a good 
approximation
for large scales (where relativistic effects should matter most), 
but not necessarily for small scales, where the ``mini-universe'' argument
is less persuasive. On the other hand, the PN approach
is well-motivated by the physics of gravity in small scales, but is
not assured to work for large scales. That the two approaches
coincide shows that, at least in some limited sense, the equations
of the SC/PN approach should give a good description of the
gravitational interactions on scales smaller than the Hubble radius.

\subsection{Pseudo-newtonian cosmology}

In PN cosmology, particles
in a comoving grid attract each other gravitationally
with a Newtonian potential. The positions of the
particles in the grid are the perturbed variables.
Although obviously limited, this approach can be used for
any configuration, not only the spherically symmetric ones.
But in order to bring the PN approach closer to the SC model, we
will adopt the same basic assumptions of the SC model for
the PN cosmological perturbations.

We consider an admixture of two fluids, cold dark matter and
dark energy. The key assumptions of the SC model (see the next subsection)
are that the density of each fluid is homogeneous at all times
in the spherical region (this is the top-hat density profile), 
and that the velocity profile preserves this homogeneity. 

The comoving coordinates are $\vec{x}_{0}=\vec{r}_{0}/a$, where
$\vec{r}_{0}$ is the homogeneous (unperturbed) physical distance
-- here, the radius of a spherically symmetric region. 
Under the assumption of the SC model, the perturbed physical 
distance (physical radius) can be written as:
\begin{equation}
\vec{r}=\left[ a\left( t \right) + f \left(t,\vec{x}_{0} \right) \right] \vec{x}_{0}\;,
\end{equation}
where $a$ is the usual scale factor and $f$ is the function that accounts
for the deviations from the background evolution. The physical velocity
is then given by:
\begin{equation}
\vec{u}=\frac{d\vec{r}}{dt}=
\left(\dot{a}+\dot{f} \right)\vec{x}_{0}=
\left(H+\frac{\dot{f}}{a} \right)\vec{r}_{0}\;,
\label{vel}
\end{equation}
where $\dot{ } = \partial/\partial t $ and 
$H=\dot{a}/a$ is the Hubble parameter. From the last equality
we can define an effective rate of expansion for the spherical region:
\begin{equation}
h=H+\frac{\dot{f}}{a} .
\end{equation}
Since the perturbed velocity is related to the peculiar velocity $\vec{v}$ by
\begin{equation}
\vec{u}=\dot{a} \, \vec{x}_{0}+\vec{v}\;,
\end{equation}
we obtain from Eq. (\ref{vel}) that:
\begin{equation}
\vec{v}= \dot{f} \, \vec{x}_{0} \; .
\end{equation}
In particular, the divergence of this velocity field is given by: 
\begin{equation}
\label{theta}
\theta \equiv \vec{\nabla}\cdot\vec{v} =
3\dot{f}+\vec{x}_{0}\cdot\vec{\nabla}\dot{f}\; .
\end{equation}
But for a top-hat profile the last term vanishes, and we obtain a 
simple relation between the local expansion rate $h$ and the 
background expansion rate $H$:
\begin{equation}
h=H+\frac{\dot{f}}{a} = H + \frac{\theta}{3a}\;.
\label{h-sc}
\end{equation}

The PN cosmological model is described by the
equations \cite{Lima:1996at}:
\begin{equation}
\frac{\partial\rho _j}{\partial t}+\vec{\nabla}\cdot\left(\vec{u}_j 
\, \rho_j\right)+p_j\vec{\nabla}\cdot\vec{u}_j=0\;,
\label{cont-pnc}
\end{equation}
\begin{equation}
\frac{\partial\vec{u}_j}{\partial t}+\left(\vec{u}_j\cdot\vec{\nabla}\right)\vec{u}_j=-\vec{\nabla}\Phi-\frac{\vec{\nabla}p_j}{\rho _j+p_j}\;,
\label{euler-pnc}
\end{equation}
\begin{equation}
\nabla^{2}\Phi=4\pi G\sum_k\left(\rho_k+3p_k\right)\;,\label{poison-pnc}
\end{equation}
where $\rho_j$, $p_j$ and $\vec{u}_j$  denote, respectively, the density,
pressure, velocity of a given cosmic fluid and $\Phi$ is the
Newtonian gravitational potential due to all the components; the equations are written in physical coordinates. 
The corresponding perturbations above the background are denoted by 
 $\delta \rho_j$, $\delta p_j$, $\vec{v}_j$ and $\phi$. 
These equations are, respectively, generalizations for fluids with 
pressure of the continuity equation, of 
the Euler equation (both valid for each fluid species $j$), and of 
the Poisson equation. 
Notice the absence
of an equation that dictates the evolution of pressure: in this
hydrodynamical approach, pressure is a thermodynamical function of the
energy, temperature, etc.

For cold dark matter and baryons the pressure is zero, but for dark energy
there is a homogeneous as well as an inhomogeneous pressure. The homogeneous
pressure is usually described in terms of a parametrized equation of state $w_e(t)$,
such that $p_e(t) = w_e(t) \rho_e(t)$. As for the pressure perturbations, we
have chosen to specify another free function, the effective sound
speed $c_{{\rm eff}}^{2}$, so $\delta p_e = c_{{\rm eff}}^{2} \delta\rho_e$.
Within the SC description, this means
that we consider an effective equation of state $w_c$ inside the
spherical region which is not necessarily equal to the background
equation of state.

With the assumptions of the SC model, the equations of PN cosmology assume
a simple form. Using the density contrast 
$\delta_j \equiv \delta\rho_j/\rho_j$ we obtain, after some algebra:
\begin{equation}
\dot{\delta}_{j}
+ 3H\left(c_{{\rm eff}\; j}^{2}-w_{j}\right)\delta_{j} 
+\frac{\theta_j}{a} \left[ 
1+w_{j}+\left(1+c_{{\rm eff}\; j}^{2}\right)\delta_{j}
\right]
=0\;,\label{cont_PN}
\end{equation}
\begin{equation}
\dot{\theta_j}+H\theta_j+\frac{\theta_j^{2}}{3a} 
% -\frac{c_{{\rm eff}\; j}^{2}}{a} \vec{\nabla} \cdot \left[\frac{\vec{\nabla} \delta_{j}}{B_j}\right]  =-\frac{\nabla^{2}\phi}{a}
=-4\pi Ga\sum_{k}\rho_{0\; k}\delta_{k}\left(1+3c_{{\rm eff}\; k}^{2}\right) \;.
\label{euler_PN}
\end{equation} 
Eq. (\ref{cont_PN}) follows from the continuity equation, and Eq. (\ref{euler_PN})
is the divergence of the Euler equation. The last equality in Eq. (\ref{euler_PN})
is found by using the Poisson equation. Note that, in general, we have separate
Euler equations for each fluid \cite{Abramo:2007iu}, but for a top-hat
profile ($ \vec{\nabla} \delta_{j} = 0$) they turn out to be identical, so there is only one $\theta$. 
The reason for that is obvious: in order to preserve the top-hat profile, all 
fluids must flow in the same way. 
Hence, in this approximation we have something similar to an effective single fluid description \cite{Avelino:2008cu}.

\subsection{The spherical collapse model}

Let us now briefly review the spherical collapse model.
This formalism describes a spherically symmetric region 
of uniform energy density $\rho_c = \rho_0+\delta\rho$ immersed in a
homogeneous universe of energy density $\rho_0$. This
spherical region will detach from the expansion
of the Universe and eventually collapse.

Consider the continuity equation
for each fluid denoted by an index $j$ in the spherical region:
\begin{equation}
\dot{\rho}_{c_{j}}+3h\left(1+w_{c_{j}}\right)\rho_{c_{j}}=0\;,
\label{sc_A}
\end{equation}
where $h=\dot{r}/r$ is the local expansion rate of that region 
and $w_{c_{j}}$ denotes the equation of state in the perturbed region.
We can regard this spherical region as a Friedmann Universe with spatial
curvature \cite{Gunn:1972sv}.
The dynamics of the coordinate $r$ is then given by the second 
Friedmann equation applied to this collapsing region:
\begin{equation}
\frac{\ddot{r}}{r}=-\frac{4\pi G}{3}\sum_{j}\left(\rho_{c_{j}}+3p_{c_{j}}\right)\;.
\label{sc_B}
\end{equation}
Equations (\ref{sc_A}) and (\ref{sc_B}), which were obtained using general
relativistic arguments, are the basic equations of the SC model.
Note that there is only one dynamical equation for
the collapsing region, which is in agreement with the single 
Euler equation that we found for the velocity field 
in the PN description, Eq. (\ref{euler_PN}).

The pressure and the energy density outside the 
spherical region are related by the background equation of state, 
$p_{0_j} = w_{0_j} \rho_{0_j} $.
Inside the spherical region these quantities can be
different from their background values, so we have now 
$p_{c_j} = w_{c_j} \rho_{c_j} $ for the collapsing region. 
In order to compare the SC formalism with the
PN equations derived in the last section, we will employ here the 
same effective sound speed we used before in order to
describe the pressure perturbations. Hence, we need to express the 
equation of state $w_{c_j}$ in terms of $c_{{\rm eff \, j}}^{2}$.
Using the density contrast $\delta_{j}=\delta\rho_{j}/\rho_{0_{j}}$,
we have that:
\begin{equation}
\rho_{c_{j}}=\left(1+\delta_{j}\right)\rho_{0_{j}}\;,
\label{contr_rel}
\end{equation}
from which it follows that:
\begin{equation}
w_{c_j}=\frac{p_{c_{j}}}{\rho_{c_{j}}}=\frac{p_{0_{j}}+\delta p_{j}}{\rho_{0_{j}}+\delta\rho_{j}}=w_{j}+\left(c_{{\rm eff}\; j}^{2}-w_{j}\right)\frac{\delta_{j}}{1+\delta_{j}} \; .
\label{wc_rel}
\end{equation}
This equation relates the equation of state in the perturbed region
to the background equation of state, the effective sound speed and
the size of perturbations. It is possible that the nature of dark
energy can be significantly changed in the perturbed region,
a phenomenon we dubbed ``dark energy mutation'' \cite{Abramo:2007mv}. 
This effect is general, occurring even at the level of linear evolution, and its magnitude 
depends on the dynamical evolution of DM and DE fluctuations -- see also Refs. \cite{Nunes:2004wn,Dutta:2006pn}.

Using now Eqs. (\ref{contr_rel}) and (\ref{wc_rel}) we can
recast Eq. (\ref{sc_A}) as:
\[
\dot{\delta}_{j}+\left(3h-3H\right)\left(1+w_{j}\right)\left(1+\delta_{j}\right)
+3h\left(c_{{\rm eff}\; j}^{2}-w_{j}\right)\delta_{j}=0\;.
\]
We can eliminate $h$ using Eq. (\ref{h-sc}), with the result:
\begin{equation}
\dot{\delta}_{j}+3H\left(c_{{\rm eff}\; j}^{2}-w_{j}\right)\delta_{j}+\left[1+w_{j}+\left(1+c_{{\rm eff}\; j}^{2}\right)\delta_{j}\right]\frac{\theta}{a}=0\;.
\label{cont_SC}
\end{equation}

Now consider the dynamical equation (\ref{sc_B}). From Eq. (\ref{h-sc}) we can write:
\begin{equation}
\dot{h}=\frac{\ddot{r}}{r}-h^{2}=\dot{H}+\frac{\dot{\theta}}{3a}-H\frac{\theta}{3a}
\; ,
\end{equation}
and substituting this expression into Eq.(\ref{sc_B}) we obtain, with the
help of Eqs. (\ref{contr_rel})-(\ref{wc_rel}), that:
\begin{equation}
\dot{\theta}+H\theta+\frac{\theta^{2}}{3a}=-4\pi Ga\sum_{k}\rho_{0_{k}}\delta_{k}\left(1+3c_{{\rm eff}\; k}^{2}\right)\;.
\label{euler_SC}
\end{equation}

Equations (\ref{cont_SC}) and (\ref{euler_SC}) are identical to
Eqs. (\ref{cont_PN})-(\ref{euler_PN}). This means that both
approaches are identical. The relations 
(\ref{h-sc}) and (\ref{wc_rel}) enable us to translate
the PN variables into the SC variables, and now it becomes
clear that the two different descriptions give the same dynamics
for a top-hat perturbation where pressure gradients are absent.

\section{Linear evolution in the SC/PN approach}

Even though we showed that the PN and SC approaches are equivalent,
that still does not mean that they are correct.
Unfortunately, presently there is no fully nonlinear general treatment 
of the evolution of perturbations in General Relativity (GR).
For this reason, we will compare our linearized results with
those obtained from linearized GR.
We will compute the linear evolution
of an overdense region well inside the matter-dominated era, and will compare
the growing mode obtained in the PN/SC formalism with the
relativistic growing mode.

The first-order equations can be linearized and recast as a single, second-order
differential equation for the density contrast of each fluid species.
We will assume that there is always a dominant ($d$) and a subdominant
($s$) fluid.
Using the scale factor $a$ for the time evolution ($' = d/da$), we obtain for the
dominant species:
\begin{eqnarray}
\delta_{d}''+\frac{\delta_{d}'}{a}\left[3\Delta_d+\frac{3}{2}\left(1-w_{d}\right)\right] 
\\ \nonumber
+\frac{3\delta_{d}}{2a^{2}}
\left[\Delta_d \left(1-3w_{d}\right)
-\left(1+w_{d}\right)\left(1+3c_{{\rm eff}_{d}}^{2}\right)\right]=0\;.
\end{eqnarray}
where
\begin{equation}
\Delta_d = \left(c_{{\rm eff}\; d}^{2}-w_{d}\right) \; .
\end{equation}
For cold dark matter ($c_{\rm eff} = w = 0$) this equation reduces to
\begin{equation}
\delta_{d}''+ \frac{3}{2} \frac{\delta_{d}'}{a}
+\frac{3\delta_{d}}{2a^{2}}
=0\;,
\end{equation}
with the well-known growing solution $\delta(a) \propto a$.
Hence, when cold dark matter is dominant, which should be the case
in the linear regime, the linear evolution of its density perturbations
is the standard one. 

For the more general case, of a dominant fluid with constant equation of state 
and constant speed of sound, the solution is given by:
\begin{equation}
\delta_{d}\left(a\right)=c_{1}a^{1+3w_{d}}+c_{2}a^{-3\left(1+2c_{{\rm eff}\; d}^{2}-w_{d}\right)/2}.
\end{equation}

Turning now to the the sub-dominant fluid, its perturbations obey the equation:
\begin{eqnarray}
\delta_{s}''+\frac{\delta_{s}'}{a}\left[3\Delta_s
+\frac{3}{2}\left(1-w_{d}\right)\right]
\\ \nonumber
+\frac{3\delta_{s}}{2a^{2}}
\left[\Delta_s \left(1-3w_{d}\right)\right]
=\frac{3\delta_{d}}{2a^{2}}\left(1+w_{s}\right)\left(1+3c_{{\rm eff}_{d}}^{2}\right)\;,
\end{eqnarray}
where
\begin{equation}
\Delta_s = \left(c_{{\rm eff}\; s}^{2}-w_{s}\right) \; .
\label{Delta_s}
\end{equation}
Assuming  again that $w_s$ and $c_{{\rm eff} \, s}^2$ are constants one has the 
solution:
\begin{equation}
\delta_{s}\left(a\right)=c_{3}a^{-3\Delta_s}+c_{4}a^{\left(3w_{s}-1\right)/2}+c_{5}a^{1+3w_{d}}\;,
\end{equation}
where
\begin{eqnarray}
c_{5}&=&\frac{1}{2}c_{1}\left(1+w_{s}\right)\left(1+3c_{{\rm eff}\; d}^{2}\right)
\\ \nonumber
& & \times
\left[
\left(1+3w_{d}\right)\left(\Delta_s+
\frac{1}{2}(1+w_{d})\right)
+\frac{1}{2}\left(1-3w_d\right)\Delta_s\right]^{-1}
\end{eqnarray}
arises from a particular solution of the inhomogeneous equation.

In particular, if matter is the dominant fluid it follows that:
\begin{equation}
c_{5}=\frac{c_{1}\left(1+w_{s}\right)}{3\Delta_s+1}
\end{equation}
and the dark energy density contrast grows in the same way as the dark matter density contrast. In addition, for the case in which $c_{{\rm eff}}^{2}=w$ one has an adiabatic condition satisfied, namely  $\delta_e=(1+w_e)\delta_m$. 
In general, however, 
the perturbations have a
non-adiabatic component and the dark energy density contrast evolves as:
\begin{equation}
\delta_{s}\left(a\right)=\frac{\left(1+w_{s}\right)}{3\Delta_s+1}\delta_{d}\left(a\right)
+c_{3}a^{-3\Delta_s}\;,
\label{deltaDE_SC}
\end{equation}
where the last term in the right-hand-side is a decreasing mode in most cases.

\section{Linear Evolution in GR}

In a previous paper \cite{Abramo:2007iu} we showed that, for a single
perfect fluid with no pressure gradients, the growing modes in the
linearized SC/PN approach coincide with those found with General Relativity (GR). 
Now we want to compare the PN and the GR solutions
for the dark energy perturbations in the linear regime, including pressure gradients.
We will consider these perturbations during the matter-dominated 
period, i.e., while DE is subdominant. This is motivated by the fact
that most observed structures were formed well into the matter-dominated
period, 
%and relativistic effects are likely to play a comparatively
%bigger role during the linear regime, when these fluctuations
%are still spread over large distances.

We consider scalar perturbations to the metric in the newtonian gauge without anisotropic stress:
\be
ds^{2}=\left(1+2\phi\right)dt^{2}-a^{2}\left(1-2\phi\right)d\vec{x}^{2}\;.
\ee
The $(00)$ and $(ii)$ componentes of Einstein equations in Fourier space are:
\be
\frac{k^{2}}{a^{2}}\phi+3H\left(\dot{\phi}+H\phi\right)=-4\pi G\sum_{j}\delta\rho_{j}\;,
\ee
\be
\ddot{\phi}+4H\dot{\phi}+\left(2\frac{\ddot{a}}{a}+H^{2}\right)\phi=4\pi G\sum_{j}\delta p_{j}\;,
\label{PhiGR}
\ee
and the conservation equations $T_{0;\mu}^{\mu}=0$ and $T_{i;\mu}^{\mu}=0$ yields:
\be
\dot{\delta}_{j}+3H\left(c_{{\rm eff}\; j}^{2}-w_{j}\right)\delta_{j}+
\left(1+w_{j}\right)\left(\frac{\theta_{j}}{a}-3\dot{\phi}\right)=0\;,
\label{deltaGR}
\ee
\be
\dot{\theta}_{j}+H\left(1-3c_{s j}^2\right)\theta_{j}
-\frac{k^{2}\delta p_{j}}{\left(1+w_{j}\right)\rho_{j}a}-\frac{k^{2}}{a}\phi=0\;,
\label{thetaGR1}
\ee
where $c_{s j}^2 = \dot{p}_j/\dot{\rho}_j$ is the adiabatic speed of sound.
. 
% 
% 
% \be
% \dot{\theta}_{j}+H\left(1-3w_{j}\right)\theta_{j}+\frac{\dot{w}_{j}}{1+w_{j}}\theta_{j}+\frac{k^{2}\delta p_{j}}{\left(1+w_{j}\right)\rho_{j}a}+\frac{k^{2}}{a}\phi=0\;.
% \label{thetaGR}
% \ee
%
%In the matter dominated regime, $\phi = {\rm const.}$ is a solution of Eq.(\ref{PhiGR}). 
%In this case, it is interesting to notice that Eq.(\ref{deltaGR}) is exactly equal to the linearized form of Eq.(\ref{cont_SC}). 
%However, Eq.(\ref{thetaGR}) coincides with
%Eq.(\ref{euler_SC}) only in the case of non-relativistic matter ($c_s = w = 0$).

In summary, the evolution of perturbations in a system consisting of dark energy and dark matter
in linearized GR is described by the following set of 5 coupled differential equations:
\begin{equation}
\ddot{\phi}+4H\dot{\phi}+\left(2\frac{\ddot{a}}{a}+H^{2}\right)\phi=\frac{3}{2}H^{2}\Omega_{e}c_{{\rm eff}}^{2}\delta_{e}\;,
\label{00_de_dm}
\end{equation}
\begin{equation}
\dot{\delta}_{m}+\frac{\theta_{m}}{a}-3\dot{\phi}=0\;,
\end{equation}
\begin{equation}
\dot{\delta}_{e}+\left(1+w_{e}\right)\left(\frac{\theta_{e}}{a}-3\dot{\phi}\right)+3H\left(c_{{\rm eff}}^{2}-w_{e}\right)\delta_{e}=0\;,
\label{cont_de_gr}
\end{equation}
\begin{equation}
\dot{\theta}_{m}+H\theta_{m}-\frac{k^{2}}{a}\phi=0\;,
\end{equation}
\begin{equation}
\dot{\theta}_{e}+H\left(1-3c_{s \, e}^2\right)\theta_{e}-\frac{k^{2}c_{{\rm eff}}^{2}\delta_{e}}{\left(1+w_{e}\right)a}-\frac{k^{2}}{a}\phi=0\;.
\label{euler_de_gr}
\end{equation}

\section{Comparison between GR and PN}

In PN cosmology the linear evolution of DM and DE is determined by the system of equations that arise from Eqs. (\ref{cont-pnc}) and (\ref{euler-pnc}) for each fluid, namely:
\begin{equation}
\dot{\delta}_{m}+\frac{\theta_{m}}{a}=0\;,
\label{cont_mat_pn_lin}
\end{equation}
\begin{equation}
\dot{\delta}_{e}+\left(1+w_{e}\right)\frac{\theta_{e}}{a}+3H\left(c_{{\rm eff}}^{2}-w_{e}\right)\delta_{e}=0\;,
\label{cont_de_st_pn}
\end{equation}
\begin{equation}
\dot{\theta}_{m}+H\theta_{m}-\frac{k^{2}}{a}\phi=0\;,
\end{equation}
\begin{equation}
\dot{\theta}_{e}+H\theta_{e}-\frac{k^{2}c_{{\rm eff}}^{2}\delta_{e}}{\left(1+w_{e}\right)a}-\frac{k^{2}}{a}\phi=0\;.
\label{euler_de_pn}
\end{equation}

Notice the absence of a dynamical equation for $\phi$. To eliminate the $k^2\phi$ terms we can use the constraint implied by the
Poisson equation in PN cosmology, Eq.(\ref{poison-pnc}). Then the time variation of the potential is determined 
by the evolution of the density constrasts. 
Also notice that these equations lack some terms when compared with their relativistic counterparts, 
as already pointed out in Ref. \cite{Mainini:2008up}. 
However, as we will show, during the matter-dominated era and on small scales, 
this discrepancy changes only the velocity that DE perturbations decay but do not modify its late-time behaviour.

In the matter-dominated regime, $\phi = {\rm const.}$ is a solution of Eq.(\ref{PhiGR}), which also arises from the system of Eqs.(\ref{cont_mat_pn_lin})-(\ref{euler_de_pn}). In this case, it is interesting to notice that Eq.(\ref{cont_de_gr}) becomes identical to 
Eq.(\ref{cont_de_st_pn}). However, Eq.(\ref{euler_de_gr}) coincides with
Eq.(\ref{euler_de_pn}) only in the case $c_{s \, e}=0$.

As we see, the equations for the growth of perturbations are different in GR and PN already in the linear regime. Now we perform a quantitative study of this difference. We will work out the case of a matter-dominated universe with a small DE component, as expected in the linear regime, in which case $\phi$ is a constant. Furthermore, to avoid further complications, we assume constant values for  $w$ and $c_{{\rm eff}}^{2}$.

Under these conditions we can write a second order differential equation for 
the linear growth of the dark energy density perturbation as
a function of the scale factor, $\delta_{e}(a)$:
\begin{equation}
\delta_{e}''+\alpha\frac{\delta_{e}'}{a}+\left[\beta+\frac{k^{2}c_{{\rm eff}}^{2}}{a^{2}H^{2}}\right]\frac{\delta_{e}}{a^{2}}=-\left(1+w\right)\frac{k^{2}}{a^{2}H^{2}}\frac{\phi}{a^{2}}\;.
\label{de-lin}
\end{equation}
This equation arises both in GR and PN: in the latter case, we keep the pressure gradient in the Euler equation (\ref{euler-pnc}), which was dropped in the case
of a top-hat perturbation. Only the parameters $\alpha$ and $\beta$ are 
different in the two cases:
\begin{eqnarray}
\alpha_{GR}&=&\frac{3}{2}+3\Delta-3w \;; \;\;  
\beta_{GR}=3\Delta\left(\frac{1}{2}-3w\right) 
\label{coe_GR}
\end{eqnarray}
\begin{eqnarray}
\alpha_{PN}&=&\frac{3}{2}+3\Delta \;; \;\;  
\beta_{PN}=\frac{3}{2}\Delta\; ,
\label{coe_PN}
\end{eqnarray}
where $\Delta$ was defined in Eq. (\ref{Delta_s}).

We will make a comparison focusing on small scales, where the PN approximation is supposed to be more accurate.
In this case, we can neglect the $\beta$ term in the square brackets of Eq.(\ref{de-lin}), and we immediately 
write a constant particular solution:
\begin{equation}
\delta_{e}=-\frac{\left(1+w\right)}{c_{{\rm eff}}^{2}}\phi.
\label{de-att}
\end{equation}

In order to solve the homogeneous equation we perform the following change
of variables:
\be
\delta_{e}\left(a\right)=x^{1-\alpha}y\left(x\right)\;,
\ee
where $x$ is defined in terms of the conformal time $\eta$ as $x = kc_{{\rm eff}}\eta$. Then Eq.(\ref{de-lin}) becomes:
\be
\frac{d^{2}y}{dx^{2}}+\frac{1}{x}\frac{dy}{dx}+\left[1-\frac{\mu^{2}}{x^{2}}\right]y=0\;,
\ee
where, according to the different coefficients in Eqs. (\ref{coe_GR})-(\ref{coe_PN}), $\mu$ assumes different values:
\be
\mu_{GR}=\pm\frac{1}{2}\left(1-6c_{{\rm eff}}^{2}\right)\;,
\ee
\be
\mu_{PN}=\pm\frac{1}{2}\left(1+6w -6c_{{\rm eff}}^{2}\right)\;.
\ee

The solutions are Bessel functions of first kind $J_{\pm\mu}\left(x\right)$.
The dark energy density contrast behaves as:
\be
\delta_{e}\left(x\right)=x^{1-\alpha}J_{\pm\mu}\left(x\right)-\frac{\left(1+w\right)}{c_{{\rm eff}}^{2}}\phi.
\label{decay}
\ee

These solutions both have an oscillatory behaviour with a decreasing amplitude proportional to $x^{1-\alpha-1/2}$ 
and they eventually reach the constant value $\delta_{e} = -\left(1+w\right)\phi/c_{{\rm eff}}^{2}$.

In order to check this analytical behaviour we numerically solve the complete system of coupled differential equations
(\ref{00_de_dm})-(\ref{euler_de_gr}). We used as illustration $c_{{\rm eff}}^{2}=-w_{e}=0.8$, 
$\Omega_{de}^{(0)} = 1-\Omega_m^{(0)} = 0.75$, and we
evolved the equations from an initial redshift of $z_i = 100$.
We examined the mode $k=100 H_0 = 0.0236h\rm{Mpc}^{-1}$, corresponding to
a physical scale of $\lambda = 266 h^{-1}\rm{Mpc}$,
well inside Hubble radius at $z_i$ and large enough to be in the
linear part of the matter power spectrum. 
As initial conditions we chose $\phi_i = -10^{-4}$, $\dot{\phi_i} = 0$ and:
\be 
\delta_m(z_i) = -2 \phi_i \left[1+ \frac{k^2 (1+z_i)^2}{3 H(z_i)^2} \right]\;; \;\;
\delta_{e}(z_i) = (1+w)  \delta_m(z_i)\;;
\ee
\be
\theta_m(z_i) = \frac{2 (1+z_i) k^2}{3 H(z_i)} \phi_i\;; \;\; \theta_e(z_i) = 0\;,
\ee
which are consistent with Einstein's equation and adiabaticity.
The result is presented in Fig. (1) and compared to the decay factor
and the final value given by Eq.(\ref{de-att}).

%%%%%%%%%%%%%%%%%%%%%%%
\begin{figure}[h,t,b]
\begin{centering}
\includegraphics[clip,scale=0.8]{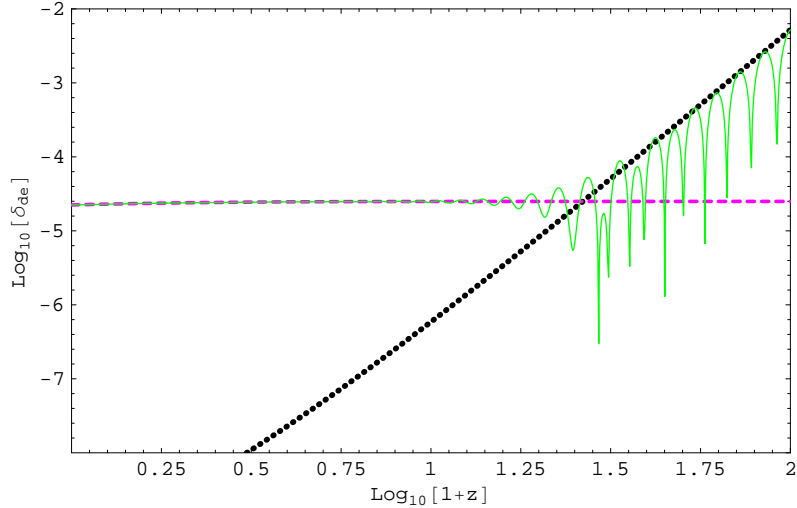}
\caption{\small \sf Linear evolution of dark energy perturbations
in GR at small scales ($k=0.0236h\rm{Mpc}^{-1}$) for $c_{{\rm eff}}^{2}=-w_{e}=0.8$.
The solid line is the solution of the complete set of 5 coupled differential
equations. The dotted line is the decay factor according to  Eq.(\ref{decay}).
The dashed line is the particular solution Eq. (\ref{de-att}).}
\par\end{centering}
\label{gr_fig}
\end{figure}
%%%%%%%%%%%%%%%%%%%%%%%%%%%%%%%%%%%%%%%

We also perform the same exercise for the PN approximation. The numerical solution of the system of Eqs. (\ref{cont_mat_pn_lin})-(\ref{euler_de_pn}) with the same parameters and initial conditions is presented in Fig. (2) 
and compared to the decay factor
and the final value given by Eq.(\ref{de-att}).  
Again we see that the qualitative analytical behaviour is reproduced by the numerical solution.

%%%%%%%%%%%%%%%%%%%%%%%
\begin{figure}[h,t,b]
\begin{centering}
\includegraphics[clip,scale=0.8]{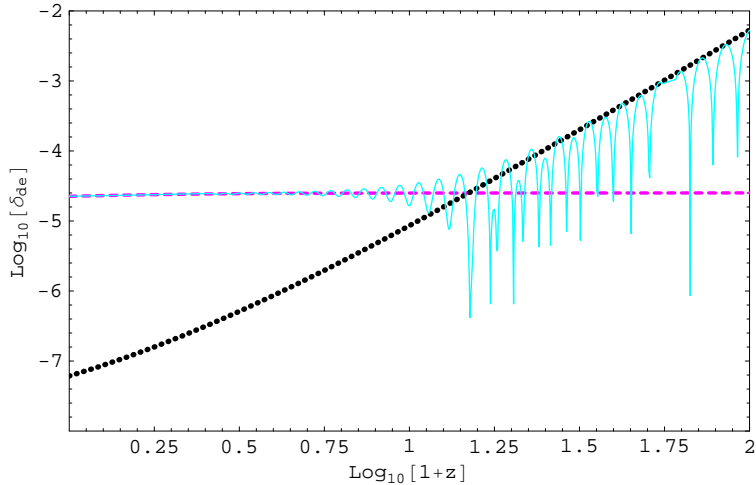}
\caption{\small \sf Linear evolution of dark energy perturbations
in the PN approximation at small scales $k=0.0236h\rm{Mpc}^{-1}$ for $c_{{\rm eff}}^{2}=-w_{e}=0.8$.
The solid line is the solution of the complete set of 4 coupled differential
equations. The dotted line is the decay factor according to  Eq.(\ref{decay}).
The dashed line is the particular solution Eq. (\ref{de-att}).}
\par\end{centering}
\label{pn_fig}
\end{figure}
%%%%%%%%%%%%%%%%%%%%%%%%%%%%%%%%%%%%%%%

Therefore, even though the equations from GR and PN are not the same 
already at the linear level, the results are not qualitatively different. 
In particular, both approaches predict the same
asymptotic behaviour for the DE perturbation. Because the
decay rate of the transient is slightly different in each case, the
time when the asymptotic regime is reached differs -- in the PN approach
this happens at a later time.

At this point we should call attention to the origin 
of an apparent discrepancy between the results obtained in this Section, namely,
a constant behaviour of the DE perturbations, Eq. (\ref{de-att}), 
and the result obtained
in Section 3, where we showed that in the SC/PN approach the DE perturbations grow as 
DM perturbations, Eq. (\ref{deltaDE_SC}). 
The reason is that in Section 3 we assumed a top-hat
profile of the perturbation, which amounts to setting $kc_{\rm{eff}}=0$ 
in the square brackets of Eq. (\ref{de-lin}). In this case, the particular solution is
\be
\delta_e(a) = -\frac{1+w}{\beta} \frac{k^2 \phi}{H^2 a^2} = \frac{3 (1+w)}{2 \beta} \delta_m(a)\;,
\label{sol_gr_c0}
\ee
where in the last equality we used Poisson's equation for the case of a dominant dark matter component, $k^2 \phi = -(3/2) H^2 a^2 \delta_m$. 
Hence, we see that indeed in this case, or in fact for perturbations with a small mode number $k$, the perturbations in DE grow at the same pace as the DM perturbations 
in the linear regime.

%Strictly speaking, this behaviour can occour only for DE models with $c_{\rm{eff}}=0$. This solution is also approximately valid in large %scales that are still inside Hubble radius.
%Hence, we see that indeed in these cases, the perturbations in DE do grow as the DM perturbations in the linear regime.

On super-Hubble scales, PN cosmology is not expected to be valid,
due to its inherently instanteneous interactions: indeed, in that 
framework perturbations with scales larger than the Hubble radius would 
behave in the same way as those well inside it. However, since we
are only interested in the evolution of perturbation which are
initially in the linear regime {\it and} well inside the Hubble radius,
this mismatch is irrelevant.
Therefore, we do not compare the PN perturbations
with the GR perturbations in large scales. 

%On super-Hubble scales, we must set 
%$k=0$ in Eq.(\ref{de-lin}), then we have solution:
%\be
%\delta_e = c_{e1}\, a^{-3(c^2_{\rm{eff}}-w)}+ c_{e2}\, a^{(6w-1)/2} \,
%\label{sol_de_large}
%\ee
%which can only grow if the effective speed of sound is very negative, %$c^2_{\rm{eff}}<w$. 

As a final remark, we recall that the analytic solution, Eq.(\ref{decay}),  
is valid only for linear perturbations during the matter-dominated period. 
In this regime the matter density constrast grows as $\delta _m \propto a$ and $\phi$ is constant in time. When the structures enter the nonlinear regime, matter fluctuations must grow faster, i.e, $\delta _m \propto a^n$, with $n>1$, then the gravitational potential should grow in time. 
Hence DE behaviour in nonlinear structures is expected to be different from the linear analysis results. However, the asymptotic constant solution in Eq.(\ref{decay}) is valid during 
the initial nonlinear process of matter collapse 
and DE fluctuations can grow with $\phi$.

\section{Conclusions}
The study of perturbations in dark energy has received a great deal of attention recently. 
DE perturbations have the potential
to alter the process of large scale structure formation in the universe. 
The existence of DE perturbations can in principle be tested 
in future surveys such as the Dark Energy Survey (DES) \cite{Abbott:2005bi} 
and EUCLID \cite{Euclid:2008eu}.
These future observations may help to distinguish among different
models of DE. 

Structure formation occurs during the nonlinear stages of the evolution 
of perturbations. Unfortunately, there is no rigorous analytical description 
of this nonlinear stage in full GR.  
Aproximation methods must be used. Possibly the most trusted method is N-body
simulations, but due to its very intensive computing requirements, it is not 
practical when one wants to study different models.
Furthermore, N-body simulations employ newtonian physics and do not
allow for the possibility of DE fluctuations.

In this paper we study two different approximation schemes, namely the Spherical Collapse and Pseudo-Newtonian
approaches. The advantage of these schemes is that DE can be fully characterized by 2 functions: 
the equation-of-state parameter $w(z)$ and
the effective speed of sound $c_{{\rm eff}}(z)$.
We show that, under a minimal set of assumptions, it is possible to \
translate one approach into the other, rendering them completely equivalent.
In order to compare these approximations with GR, we study perturbations in 
the linearized regime with all approaches.
When the assumptions about the pressure perturbations are the same both in 
GR and PN/SC we find that the fluctuactions present the same qualitative 
behaviour, lending support to the approximations. 
However, in order to establish more firmly the validity of the 
approximations in the nonlinear regime, 
a comparison should be made with some nonperturbative model in GR, 
such as an extended LTB class of models, including fluids with pressure. 
Work along this direction is in progress.

%When one considers pressure gradients in DE component its linear evolution is quite different from the one 
%expected in SC type models. This is due to different model construction: in SC models an overdense or 
%underdense region evolves like a local homogeneous universe, without any density or pressure gradients. 
%This is a fairly good approximation for DM evolution, at least to determine its halo abundance, but 
%debatable for DE. Although note that only DE models with negative pressure gradients are ruled out by 
%observations \cite{Sandvik:2002jz}. 

\section*{References}

\bibliographystyle{h-physrev3}
\bibliography{Equivalence}

%\end{thebibliography}

\end{document}